\documentclass[aps,pra,twocolumn]{revtex4-1}

\usepackage{amsmath}
\usepackage{mathrsfs}
\usepackage{amssymb}
\usepackage{amsfonts}
\usepackage{dsfont}
\usepackage{IEEEtrantools}
\usepackage{bm}
\usepackage{amsthm}
\usepackage{stmaryrd}

\renewcommand{\H}{\mathcal{H}}
\newcommand{\ket}[1]{|#1 \rangle}

\newcommand{\ketbra}[2]{|#1 \rangle\langle #2|}

\newcommand{\proj}[1]{\ketbra{#1}{#1}}

\newcommand{\da}{\downarrow}
\newcommand{\ua}{\uparrow}
\newcommand{\ok}{\mathtt{ok}}

\newcommand{\AND}{\mathtt{AND}}
\newcommand{\OR}{\mathtt{OR}}
\newcommand{\id}{\mathds{1}}
\DeclareMathOperator{\tr}{Tr}

\usepackage{csquotes}
\usepackage{enumerate}
\usepackage{pgfplots}
\usepackage{tikz}
\usetikzlibrary{calc,positioning}

\usepackage{hyperref}

\theoremstyle{definition}

\newtheorem{protocol}{Protocol}
\theoremstyle{remark}

\begin{document}

\title{The measurement problem is the measurement problem is the measurement problem} 
\author{Veronika Baumann, Arne Hansen, and Stefan Wolf}
\affiliation{Facolt\`a di Informatica, 
Universit\`a della Svizzera italiana, Via G. Buffi 13, 6900 Lugano, Switzerland}

\date{\today}

\begin{abstract}
\noindent
Recently, it has been stated that {\em single-world interpretations of quantum theory are logically inconsistent\/}. 
The claim is derived from contradicting statements of agents in a setup combining two Wigner's-friend experiments.
Those statements stem from applying the measurement-update rule subjectively, i.e., only for the respective agent's own measurement.
We argue that the contradiction expresses the {\em incompatibility of collapse and unitarity\/} --- resulting in different formal descriptions of a measurement --- and does not allow to dismiss any specific interpretation of quantum theory. 
\end{abstract}

\maketitle

\section{Introduction}
\noindent

The mathematical formalism of quantum theory has been celebrated for its success; it has been well-tested and, so far, not been falsified~\cite{Popper1934} --- even in cases where its predictions are counter-intuitive and paradoxical \cite{aspect1982experimental,bouwmeester1997experimental,stefanov2002quantum,ma2012experimental,denkmayr2014observation}. 
Nevertheless, there are ongoing controversies: On the one hand, there are various attempts to resolve the conflict of the apparent collapse during a measurement with the unitarity evolution. 
On the other hand, controversial discussions are led on how to understand and to interpret the formalisms. 

The authors of~\cite{FrRen} claim to have proven that \emph{single-world interpretations of quantum theory cannot be self-consistent}. 
They consider a combination of two Wigner's-friend \emph{gedankenexperiments} and derive a contradiction regarding the statements of the agents involved. 
The contradiction is, actually, a result of the \emph{quantum-mechanical description} of the measurement:
The agents in the setup attribute a collapse \footnote{The term collapse (of the wave function) is to be understood as the applicability of the measurement update rule throughout this paper.} merely to their \emph{own} measurement. 
We subsequently refer to this as \emph{subjective collapse}. 
It is, however, subjective only in the case of encapsulated observers where unitarity and collapse are conflicting, see~\ref{sec:Wigner} and~\ref{sec:ext_Wigner}. 
Agents measuring the same quantum system --- not each other's memories --- can objectively agree on a collapse occurring in every measurement.

In our opinion, the subjective-collapse model, as Everett's relative-state model~\cite{everett1957relative}, or other collapse models such as GRW~\cite{GRW85}, is a mathematical formalism and not an interpretation.
All the above formalisms yield the same predictions on experiments until one considers encapsulated observers; and then --- if ever possible --- a Wigner's-friend experiment will have to decide among them.

In~\cite{FrRen}, the authors introduce a framework of so-called \emph{stories} and \emph{plots} that captures ``what could be said'' --- in particular about entities in a physical experiment;
the contradiction in question is obtained in this framework.
We argue that the use of the framework to rule out certain interpretations of quantum theory as done in~\cite{FrRen} is debatable.
This doubt has been supported recently~\cite{sudbery2016single} by a consistent description of the setup in terms of a generalised version of Bohmian mechanics. 

We first discuss the \emph{story-plot framework} and hope to clarify how it relates to classical information. 
We critically examine the notion of ``many worlds'' established in the framework.
Furthermore, we examine the original Wigner's-friend experiment and formulate a contradiction analogous to that in~\cite{FrRen} in terms of stories and plots.
Said contradiction, therefore, can be seen to reflect the measurement problem.
Finally, we consider the setup in~\cite{FrRen} and explicitly show how the contradiction arises from the subjective collapse.
 
\section{The Story-Plot Framework}
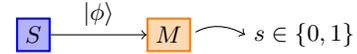
\begin{figure}
\centering
\begin{tikzpicture}[scale=0.6]
  \node[fill=blue!30, draw=blue, thick] (s) at (0,0) {$S$};
  \node[fill=orange!30, draw=orange, thick] (m) at (3,0) {$M$};
  \node[] (r) at (6,0) {$s\in\left\{0,1\right\}$};
  \draw[->] (s) -- node [midway,above ] {$\ket{\phi}$} (m);
  \draw[->,shorten <=2pt] (m.east) to[bend left] (r.west);
\end{tikzpicture}
\caption{A source~$S$ emits a 2-dimensional quantum state $\ket{\phi}$ that is measured (in~$M$) to yield some binary result~$s$.}
\label{fig:Stern-Gerlach}
\end{figure}

\noindent
The framework addresses the question what can possibly be said about a physical experiment. 
A simple example from quantum physics is the Stern-Gerlach setup, depicted in Fig~\ref{fig:Stern-Gerlach}.
After recapitulating the framework in~\ref{ssec:stories_events_plots}, we discuss in~\ref{ssec:critique} how the compatibility constraint in~\cite{FrRen} relates to \emph{interoperability} --- the ability to copy information --- and how their definition of ``many worlds'' is in conflict with the definiteness of measurement results. 
The latter relates to the \emph{distinguishability} of the information obtained by the measurement.
Finally, in~\ref{ssec:inf_qm}, we argue that the set of possible stories about a quantum-mechanical experiment can be formally represented by joint, conditional probability distributions.
\subsection{Events, Stories, and Plots}
\label{ssec:stories_events_plots}
\noindent
In~\cite{FrRen}, a story~$s$ is defined as ``an account of events that occur.''
Neither of the terms ``event,'' ``occur,'' and ``real'' is unambiguous in this context, 
in particular if there is no further specification of the theoretical foundation.
We will generalize the concept slightly and take a story to be a priori ``anything that can be stated'' \footnote{Any further constraints on what information the stories are supposed to contain can be imposed with an appropriate choice of event sets.}.
Statements (or accounts or stories) can be encoded in a finite bit string.
These bit strings carry classical information that is preserved when copied --- we will term this feature \emph{interoperability} --- and is based on a notion of \emph{distinguishability}, i.e., different stories can be told apart from one another.
Thus, stories are entities of classical information, even though we might talk about quantum systems.

A possible story is for instance
\begin{displayquote}
``The source emits a photon in the state $\ket{\phi} = \alpha \ket{0} + \beta\ket{1}$ which is measured in the basis $\left\{ \ket{0}, \ket{1} \right\}$ and yields the outcome~$0$.''
\end{displayquote}
Another example is 
\begin{displayquote}
``If a source emits a photon in the state $\ket{\phi} = \alpha \ket{0} + \beta\ket{1}$, a measurement in the basis $\left\{ \ket{0}, \ket{1} \right\}$ \emph{might} yield the outcome~$0$ or the outcome~$1$.''
\end{displayquote}
While the first has a \emph{factual} character, the second accounts for different \emph{possible} outcomes.
Both are elements in the set of all possible stories, $\Sigma$.
If the setting of the story (e.g., the experimental setup) is known, and the content has enough structure, by interoperability, one can represent it more concisely with a set of ordered tuples of parameters --- each tuple capturing one of the instances of information. 
From now on we assume --- as in~\cite{FrRen} --- that stories have some temporal structure. 
Thus, the first parameter in the tuple is a time index from a discrete ordered set $\mathcal{T}$ allowing for a notion of \emph{before} and \emph{after}. 
All other information is stored in a tuple $\bm x \in V$ provided it can be matched with the representation. 
As the story is finite, it can represent, for instance, real numbers only up to some finite precision.
One might nonetheless take the values of the parameters embedded in a continuous space and, therefore, the sets in the Cartesian product~$V$ to be real. 
The set of all possible events given the temporal structure and some Cartesian product~$V$ is then
\begin{equation}\label{eq:eventset}
  E = \left\{ (t, \bm x) \in \mathcal{T} \times V \right\}.
\end{equation}
A \emph{deduction function} maps stories into the power set $\mathcal{P}(E)$ and assigns to each story a set of events, called a \emph{plot}:
\begin{IEEEeqnarray*}{RRCL}
  d_E : & \Sigma & \to & \mathcal{P}(E) \\
  &s & \mapsto & d_E(s) =: s^E \subset E \ .
\end{IEEEeqnarray*}
The deduction function depends on the underlying event set. 
Some information of the story~$s$ might get lost in its representation by events, $d_E(s)$.
While in~\cite{FrRen},~$d_E$ is taken to be a partial function, we rather map stories that are meaningless with respect to the event set~$E$ to the empty set.

\subsection{Critique of the Story-Plot Framework}
\label{ssec:critique}
\noindent
\paragraph{Compatibility and interoperability} By interoperability one can require different plots for different event sets stemming from the \emph{same story} to be in some sense \emph{compatible}. 
Two event sets~$E$ and~$F$ encompass the same information about a story~$s$, if the plots $d_E(s)$ and $d_F(s)$ can be related bijectively. 
If, however,~$F$ captures only a part of the information captured by~$E$, i.e., $V_E = V_F \times W$ for some set~$W$, then for any $(t,\bm x)\in d_F(s)$, there has to exist a corresponding $ (t,\bm x,\bm y) \in d_E(s)$, that is
\begin{equation}\label{eqn:campatibility1}
  (t, \bm x) \in d_F(s) \quad \Leftrightarrow \quad \exists \bm y\in W: (t, \bm x, \bm y) \in d_E(s).
\end{equation}
Generally, if two event sets~$E$ and~$F$ have some ``overlap,'' i.e., there exist sets~$W_E$ and~$W_F$ such that $V_E \times W_E = V_F \times W_F$, then the compatibility condition is 
\begin{IEEEeqnarray}{RL}\label{eqn:campatibility2}
  &\exists \bm z \in W_F: (t, \bm x,\bm z) \in d_F(s) \\
  \Leftrightarrow \quad &\exists \bm y \in W_E: (t, \bm x, \bm y) \in d_E(s).\nonumber
\end{IEEEeqnarray}
This is the compatibility constraint introduced in~\cite{FrRen} and demands different plots to be copies of the information of the same corresponding story. It is possible due to the interoperability of classical information.\\

\paragraph{Many worlds and distinguishability} Stories allow for multiple events happening at the same time, as illustrated in the following examples.
\begin{displayquote}
``Alice and Bob perform their respective measurements.''
\end{displayquote}
\begin{displayquote}
``Alice measures either~$0$ or~$1$.''
\end{displayquote}
In the first story two a priori uncorrelated events happen ``at the same time,'' by which we rather mean that we cannot distinguish which one happened before the other.
The second story refers to multiple possible results.
Both stories have plots with at least two events
\begin{equation*}
  (t, \bm{x}_1), (t,\bm{x}_2) \in d_E(s)
\end{equation*}
with the same time parameter~$t$.
While in the first story those two events are related by $\AND$ for two independent measurements, the latter contains events that are connected by $\OR$.
From the plots alone, it cannot be deduced how two events with the same time parameter are related to one another.
\begin{figure}
\begin{center}
\begin{tikzpicture}[scale=.6]
  \coordinate (p1) at (-1,0);
  \coordinate (p21) at (2,2);
  \coordinate (p22) at (2,-2);
  \coordinate (p31) at (4,3);
  \coordinate (p32) at (4,1);
  \coordinate (p33) at (4,-1);
  \coordinate (p34) at (4,-3);
  \coordinate (p41) at (5,3.5);
  \coordinate (p42) at (5,2.5);
  \coordinate (p43) at (5,1.5);
  \coordinate (p44) at (5,0.5);
  \coordinate (p45) at (5,-0.5);
  \coordinate (p46) at (5,-1.5);
  \coordinate (p47) at (5,-2.5);
  \coordinate (p48) at (5,-3.5);
  \draw[] (p1) to[bend left] (p22);
  \draw[] (p21) to[bend left] (p31);
  \draw[] (p22) to[bend left] (p33);
  \draw[] (p22) to[bend right] (p34);
  \draw[] (p31) to[bend left] (p41);
  \draw[] (p31) to[bend right] (p42);
  \draw[] (p32) to[bend right] (p44);
  \draw[] (p33) to[bend left] (p45);
  \draw[] (p33) to[bend right] (p46);
  \draw[] (p34) to[bend left] (p47);
  \draw[] (p34) to[bend right] (p48);
  \draw[line width=3pt, orange] (p1) to[bend right] (p21);
  \draw[line width=3pt, orange] (p21) to[bend right] (p32);
  \draw[line width=3pt, orange] (p32) to[bend left] (p43);
  %\node[right] (eye) at (p43) {\Large\faEye \ $\rightarrow$ \ \faCommenting};
  \draw[thick] ($(p43)+(.6,.3)$) to[bend right] ($(p43)+(1.2,.0)$);
  \draw[thick] ($(p43)+(.6,-.3)$) to[bend left] ($(p43)+(1.2,.0)$);
  \draw[thick] ($(p43)+(.65,.2)$) to[bend right] ($(p43)+(0.65,-.2)$);
  \node[] at ($(p43)+(2.5,0)$) {$\rightarrow$ story};
\end{tikzpicture}
\end{center}
\caption{Observers performing one experiment see exclusively one of the branches in the multiverse. This selection of one branch upon measurement ensures the definiteness of the measurement result.}
\label{fig:many worlds}
\end{figure}
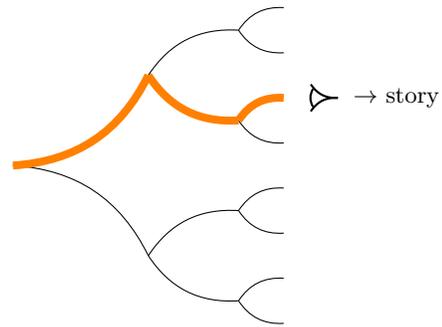
In~\cite{FrRen}, the authors allow for $\AND$-connected events in stories about \emph{one single measurement} and refer to this as \emph{many worlds}. 
This, however, is in conflict with the \emph{definiteness} of measurement outcomes, in the sense that each single instance of a measurement gives exclusively one result. The ability to infer distinguishable, classical information from it might be seen as the defining property of a measurement.
The \emph{distinguishability} of classical information --- a classical bit is either~$0$ or~$1$ --- then \emph{requires}, by definition, measurement outcomes to be definite from the point of view of a given observer.
Relating two events describing results for a single measurement with $\AND$ is then a contradiction in itself. 
The statement ``Alice measures 0 and 1'' does not describe a measurement in the above sense, since it does not describe Alice inferring a classical bit.
In our understanding, two events in a plot of a story about \emph{one measurement}, therefore, have to be connected by $\OR$. 
Connecting such events by $\AND$ is questionable.

In a many-worlds interpretation, a tree as shown in Fig~\ref{fig:many worlds} is commonly used to depict the multiverse~\cite{wallace2012emergent}.
Such a tree illustrating all \emph{possible} results --- i.e., the ``multiverse'' ---, can also be constructed for collapse models. 
The probability weights associated to the branches in the different formalisms can, however, differ. 
Whatever the interpretation, statements regarding actual measurement results of an experiment refer to \emph{one particular branch} of the multiverse. 
Within \emph{one} such branch, there are no $\AND$-related outcomes for a single measurement.

\subsection{Plots and Quantum Theory}
\label{ssec:inf_qm}
\noindent
In quantum mechanics, Born's rule relates the measurement outcomes with subspaces in a sample space $\Omega$ and a corresponding probability distribution~$P$.
By definiteness of the outcomes we assume that $\Omega$, in a single measurement, falls into a disjoint union of at least two subsets
\begin{equation*}
  \Omega = \bigcup_k V_k, \quad V_k \cap V_l = \emptyset \ \forall \ k\neq l,
\end{equation*}
where each correspond to different measurement results and subsequently to different events. 
Stating disjointness is then equivalent to saying that two results \emph{cannot occur simultaneously}, i.e., 
\begin{equation*}
  P(V_k\cap V_l) = 0 \quad \forall \ k \neq l.
\end{equation*}

Consider a quantum experiment with multiple parties measuring each part of a density matrix $\rho$ on~$\H_{W_1}\otimes \H_{W_2} \cdots$. The overall probability distribution~$P(w_1,w_2,\ldots)=\tr(\rho \pi_{w_1}\otimes \pi_{w_2} \cdots) $ encodes all possible combinations of measurement results.
Each party that knows the entire setup and thus $\rho$, can compute her own probability distribution~$P_k(w_k)= \tr_{W'}(\rho \ \id\otimes \pi_{w_k})$, where $\tr_{W'}$ is the partial trace over all other parties. The conditional probability distribution~$P(w_1,\ldots, w_{k-1}, w_{k+1}, \ldots \mid w_k)$ reflects the party's knowledge about the other results, given her own outcome. She might have either used Bayes' formula or computed $\tr_{W_k}( \rho \ \id\otimes \pi_{w_k})$ and applied Born's rule to the renormalized state. 

All plots corresponding to a quantum experiment then relate to the joint probability distribution and can only contain~$\OR$-related events referring to different results of one measurement.
 \section{Wigner's friend}
\label{sec:Wigner}

\begin{figure}
\centering
\begin{tikzpicture}[scale=0.6]
  \draw[thick] (-1,-2) rectangle (5.3,2);
  \node[fill=blue!30, draw=blue, thick] (s) at (0,0) {$S$};
  \node[fill=orange!30, draw=orange, thick] (m) at (2,0) {$M_F$};
  \node[fill=orange!30, draw=orange, thick] (M) at (8,0) {$M_W$};
  \node[fill=blue!30, draw=blue, thick, rounded corners=3pt] (r1) at (4,1) {$\ket{\ua}\ket{\text{u}}$};
  \node[fill=blue!30, draw=blue, thick, rounded corners=3pt] (r2) at (4,-1) {$\ket{\da}\ket{\text{d}}$};
  \draw[->] (s) -- node [midway,above] {$\ket{\phi}$} (m);
  \draw[->,shorten >=2pt,shorten <=2pt] (m) to[bend left] (r1);
  \draw[->,shorten >=2pt,shorten <=2pt] (m) to[bend right] (r2);
  \draw[->,thick] (5.3,0) to (M);
\end{tikzpicture}
\caption{The source~$S$ emits a quantum state (e.g., a qubit) $\ket{\phi}$, which is measured by the friend in some basis (e.g., \{$\ket{\ua}, \ket{\da}$\}). Wigner then measures the joint system of the state emitted by the source and the friend's memory.}
\label{fig:Wigner}
\end{figure}
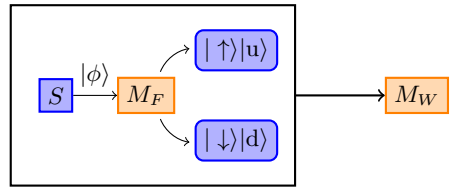
The setup of the original Wigner's-friend experiment is depicted in Fig~\ref{fig:Wigner}.
An \emph{observer} --- the friend~$F$ --- performs a measurement~$M_F$ on the quantum system emitted by the source~$S$.
Both the source and the friend constitute a joint quantum system which is then measured by a \emph{superobserver} --- Wigner~$W$ --- with a projective measurement~$M_W$.
We say, Wigner has \emph{full quantum control} over the friend's lab.\\
According to textbook quantum mechanics, the friend's measurement induces a collapse to the eigenvector associated with the observed measurement result.
To Wigner, however, the source and the friend's memory appear as one big quantum system, which supposedly evolves unitarily, i.e., without a collapse.
If the friend's measurement does not induce a collapse, does he actually obtain a definite measurement result?
Deutsch's variant of the experiment~\cite{deutsch1985quantum} addresses this issue. 
The friend raises a flag stating whether he observed a definite outcome or not.

We differentiate between three models. 
The \emph{no-collapse} model, which can be associated with Everett's relative-state formalism~\cite{everett1957relative}, removes the collapse completely.
The \emph{objective collapse} model refers to theories assuming a collapse in every measurement. 
In that case, even for Wigner the evolution of the joint system is not unitary due to the friend's measurement.
Finally, the \emph{subjective collapse} model, with each agent assuming a collapse merely in his own measurement, is labelled as \emph{standard quantum theory} in~\cite{FrRen}.  
While in the first two all agents make consistent predictions about all measurements, the latter allows Wigner and his friend to make predictions that contradict each other.
One further has to assume that the friend can actually \emph{test} his predictions about Wigner's result.
The possibility of communication between encapsulated observers is questionable.
In that sense, the subjective collapse might be sufficient if the ``wrong'' prediction cannot be tested.

\subsection{Everettian Description of the Experiment}
\label{ssec:Everett_wigners_friend}
\noindent
According to Everett, the friend's measurement is an isometry, correlating his memory state with the different elements of the measurement basis $\{\ket{m_F^{(i)}}\}_i$,
\begin{IEEEeqnarray}{RCL}\label{eqn:stdIsom}
  V_F : \quad \H_S & \to &  \H_S \otimes \H_F\\ 
  \ket{m_F^{(i)}} & \mapsto & \ket{m_F^{(i)}} \otimes \ket{z_i} \quad \forall i\nonumber , 
 \end{IEEEeqnarray}
where $\{\ket{z_i}\}_i$ is an orthogonal set in the memory system recording the result.
Consistently, we can model Wigner's measurement with an isometry
\footnote{The combined system of a qubit emitted by the source and the friend's memory state is a four dimensional quantum system. The image of the isometry~$V_F$ is, however, two-dimensional. So Wigner can choose his measurement basis such that two of the four results do not occur.}
\begin{IEEEeqnarray}{RCL}\label{eqn:VWigner}
  V_{W} : \H_S\otimes \H_F & \to & \H_S\otimes \H_F \otimes \H_W
\end{IEEEeqnarray}
to finally obtain the joint memory system as
\begin{equation*}
 \rho_{\text{mem}} = \tr_S\left( V_W \ V_F \ \proj{\psi_S} \ V_F^{\dagger} \  V_{W}^{\dagger}\right),
\end{equation*}
where $\ket{\psi_S}\in\H_S$ is the state emitted by the source.
The density matrix $\rho_{\text{mem}}$ encodes possible correlations between the memory states of Wigner and his friend.
To compute it one merely needs to know the two measurement bases, the one of the friend and the one of Wigner.
We assume that the source emits a superposition~$\ket{\psi_S} = \frac{1}{\sqrt{2}} \left( \ket{\ua} + \ket{\da} \right)$ , that is measured by the friend in the basis $ \left\{ \ket{\da}, \ket{\ua} \right\}$ before Wigner measures the joint system in a product basis $\left\{\ket{\ua} \otimes \ket{u}, \ket{\da} \otimes \ket{d}\right\}$. 
The corresponding density matrix 
\begin{IEEEeqnarray}{RL}\label{eqn:rho_mem1}
  \rho_{\text{mem}}^{(1)} = \frac{1}{2} 
  &\left( \proj{u, U} + \proj{d,D} \right)
\end{IEEEeqnarray}
shows that Wigner will know the measurement result of the friend. 
Alternatively, Wigner can measure in the superposition basis $\{\ket{\phi^{\pm}}= \sqrt{1/2} (\ket{\ua} \otimes \ket{u} \pm  \ket{\da} \otimes \ket{d}) \}$
resulting in memory state
\begin{IEEEeqnarray}{RL} \label{eqn:rho_mem2}
  \rho_{\text{mem}}^{(2)} = \frac{1}{2} 
  &\left( \proj{u} + \proj{d} \right) \otimes \proj{+},
\end{IEEEeqnarray}
where $\ket{+}$ corresponds to Wigner observing the joint system to be in state $\ket{\phi^{+}}$. 
As the density matrix is a product, Wigner cannot extract any information about the friend's measurement result.
To incorporate a collapse in this formalism, one replaces the source with the measured state. If the friend's result was~$u$, the memory state would be
\begin{IEEEeqnarray}{RL}\label{eqn:rho_clps}
  \rho_{\text{mem}}^{\text{clps}} = &\tr_S\left( V_W \ V_F \ \proj{\ua} \ V_F^{\dagger} \  V_{W}^{\dagger}\right) \\
  = & \tr_S\left( V_W \ \proj{\ua} \otimes \proj{u} \  V_W^{\dagger} \right). \nonumber
\end{IEEEeqnarray}
If Wigner then measured in the superposition basis, he would measure $\ket{\phi^-}$ with non-zero probability in case of a collapse.

\subsection{The Experiment in the Story-Plot Framework}
\begin{figure}
\centering
\begin{tikzpicture}[scale=0.6, box/.style={thick, color=blue, fill=blue!30, rounded corners=3pt}]
  \begin{scope}[shift={(-3,3)}]
    \draw[ box] (-2,-3) rectangle (2,2);
    \node at (-2.5,2.5) {$\bm{s^W}$};
    \node (W_e1) at (0,1) {$(t_1,\star, x=0)$};
    \node (W_e0) at (0,-1){$(t_2,\star,\bm0)$};
    \node (W_e2) at (0,-2) {$(t_2,\star,\bm1)$};
    \node[draw, thick, color=red, fill=red!10, circle] at (-2,-1.5) {$\Large \lightning $};
  \end{scope}
  \begin{scope}[shift={(3,3)}]
    \draw[ box] (-2,-3) rectangle (2,2);
    \node at (2.5,2.5) {$\bm{s^F}$};
    \node (F_e1) at (0,1) {$(t_1,\star, 0)$};
    \node (F_e2) at (0,-2) {$(t_2,\star, y=1)$};
  \end{scope}
  \draw[->, thick] (W_e1) to node[midway, above] {\eqref{compx1} $\;\;$} (F_e1);
  \draw[->, thick] (F_e1) to node[midway, right] {} (F_e2);
  \draw[->, thick] (F_e2) to node[midway, below] {$\;\;\;$ \eqref{compx2}} (W_e2);
\end{tikzpicture}
\caption{The contradiction in Deutsch's version of the Wigner's-friend experiment: If the friend observes a definite result, he encodes this in the bit $x=0$ and, therefore, $(t_1,\star,0) \in s^F$. 
Applying the measurement update rule --- the subjective collapse --- he deduces that Wigner should measure ``$+$'' or ``$-$'' with equal probability, $(t_2, \star, y=1) \in s^F$. 
Wigner, however, deduces  $(t_1,\star,x=0) \in s^W$ from the communicated bit (see~\eqref{compx1}). 
Assuming unitarity, he observes $\ket{\phi^+}$ in the superposition measurement, i.e., $(t_2, \star, 0) \in s^W$. 
Wigner knows from the bit~$x$ that the friend obtained a definite outcome, attributed a collapse and concluded also ``$-$'' to be possible.
By this retrace of~$F$'s conclusion together with~\eqref{compx2}, Wigner arrives at the contradiction.}
\label{fig:WF-paradox}
\end{figure}
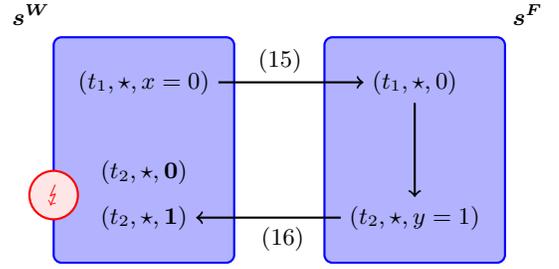
\noindent
The friend will account for his measurement with a plot
\begin{equation} \label{sfriend0}
s_0^F = \{ (t_1, z) \}
\end{equation}
containing a single event and~$z$ being the measurement result he \emph{observed}.
The underlying event set is $E = \left\{ \mathcal{T} \times \left\{0,1\right\} \right\}$.
Wigner, on the other hand, has a plot
\begin{equation} \label{swigner0}
s_0^W = \{(t_2, w) \},  
\end{equation}
where~$w$ is the outcome Wigner \emph{observed} in his measurement on the combined system. 
Compatibility constraints arise when Wigner and his friend \emph{deduce} something from their measurements. Generally, their plots will then be of the form
\begin{equation} \label{sfriend0}
s^F = \{ (t_0,\bm{e}),(t_1, z'), (t_2, \bm{y}_w )\} , 
\end{equation}
where $\bm e$ and $\bm{y}_w $ describe what the friend can \emph{deduce} about the source and Wigner's results. 
Wigner's plot will be
\begin{equation} \label{swigner0}
s^W = \{(t_0,\bm{e'}),(t_1, \bm{y}_f), (t_2, w') \},  
\end{equation}
where $\bm e'$ and $\bm{y}_f$ describe his \emph{deductions}. 
Using the same shortened notation as in~\cite{FrRen}, compatibility would then require that
\begin{IEEEeqnarray}{RLCRL}\label{compWF}
 (t_1, z',\star) &\in s^F \quad &\Leftrightarrow & \quad  (t_1, z=z', \star) & \in s^W \label{WF1st} \\
 (t_2, \star, w=w') &\in s^F \quad &\Leftrightarrow & \quad  (t_2, \star, w') & \in s^W,\label{WF2nd} 
\end{IEEEeqnarray}
which means that if one party can deduce the outcome of the other party's measurement with certainty, this should be what that party observes.
Deductions will in the following be written as equalities, while observed quantities are represented by their value.

In Deutsch's version of the Wigner's-friend experiment, the friend and Wigner communicate an additional bit~$x$, encoding whether the friend observed a definite outcome or not. 
Assume that they further have to answer the question ``\emph{Can} Wigner measure $\ket{\phi^-}$?'' encoded in another bit~$y$.
This introduces further compatibility constraints 
\begin{IEEEeqnarray}{RLCRL}\label{compWFx}
 (t_1,\star,x') &\in s^F \quad &\Leftrightarrow & \quad  (t_1,\star, x=x') & \in s^W \label{compx1} \\
 (t_2, \star, y=y') &\in s^F \quad &\Leftrightarrow & \quad  (t_2, \star, y') & \in s^W.\label{compx2} 
\end{IEEEeqnarray}
The subjective-collapse model gives rise to the contradiction depicted in Fig~\ref{fig:WF-paradox}.
 \section{The Frauchiger/Renner Protocol}
\label{sec:ext_Wigner}
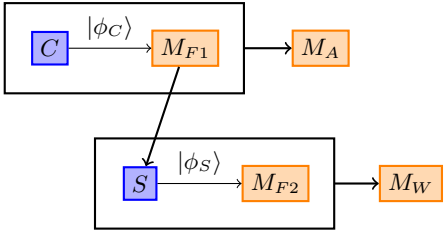
\begin{figure}
\centering
\begin{tikzpicture}[scale=0.6]
  \draw[thick] (-1,-1) rectangle (4.3,1);
  \node[fill=blue!30, draw=blue, thick] (c) at (0,0) {$C$};
  \node[fill=orange!30, draw=orange, thick] (mf1) at (3,0) {$M_{F1}$};
  \node[fill=orange!30, draw=orange, thick] (MA) at (6,0) {$M_A$};
  \draw[->] (c) -- node [pos=0.5,above] {$\ket{\phi_C}$} (mf1);
  \draw[->,thick] (4.3,0) to (MA);
  \begin{scope}[shift={(2,-3)}]
  \draw[thick] (-1,-1) rectangle (4.3,1);
  \node[fill=blue!30, draw=blue, thick] (s) at (0,0) {$S$};
  \node[fill=orange!30, draw=orange, thick] (mf2) at (3,0) {$M_{F2}$};
  \node[fill=orange!30, draw=orange, thick] (MW) at (6,0) {$M_W$};
  \draw[->] (s) -- node [pos=0.5,above] {$\ket{\phi_S}$} (mf2);
  \draw[->,thick] (4.3,0) to (MW);
  \end{scope}
  \draw[->,thick] (mf1) to (s);
\end{tikzpicture}
\caption{The modified extended Wigner setup.}
\label{fig:modWigner}
\end{figure}
\noindent
The setup introduced in~\cite{FrRen} allows to establish contradicting predictions, as for the Wigner's-friend experiment above, more strikingly and in a single-run.
It consists of two standard Wigner's-friend experiments, where the friend in the first acts as the source for the second.
The setup is given by the following protocol.
\begin{protocol}
The protocol assumes four parties, two friends,~$F_1$ and~$F_2$, and two super-observers, Wigner~$W$ and his assistant~$A$, with full quantum control over~$F_1$ and~$F_2$ respectively. 
The steps of the protocol are
\begin{enumerate}
\setcounter{enumi}{-1}
  \item At time~$t_0$, a source emits a quantum coin in the superposition $\ket{\psi_c} = \sqrt{\frac{1}{3}}\ket{h} + \sqrt{\frac{2}{3}} \ket{t}$ of head and tail. 
  \item At time~$t_1$, the first friend~$F_1$ measures the coin and prepares a spin state $\ket{\phi_s}=\ket{\da}$ if the result of the measurement was head and a spin state $\ket{\phi_s} = \sqrt{\frac{1}{2}} (\ket{\da} + \ket{\ua})$ if the result was tail.
  The measurement is given by an isometry correlating the brain state $\ket{\phi_{F1}} \in \left\{ \ket{H}, \ket{T}\right\}$ of~$F_1$ with the different coin states.
  \item At time~$t_2$, the second friend~$F_2$ measures the spin state in the basis $\left\{ \ket{\ua}, \ket{\da} \right\}$.
  The measurement is given by an isometry correlating the brain state $\ket{\phi_{F2}} \in \left\{ \ket{U}, \ket{D}\right\}$ of~$F_2$ with the different spin states.
  \item At time~$t_3$, the assistant~$A$ measures the coin~$C$ and the brain state of~$F_1$ in a basis 
    \begin{IEEEeqnarray*}{RL}
      \big\{ &\ket{o} = \sqrt{1/2} ( \ket{h,H} - \ket{t,T} ),\\
      &\ket{f} = \sqrt{1/2} ( \ket{h,H} + \ket{t,T} ) \big\}.
    \end{IEEEeqnarray*}
  \item At time~$t_4$, Wigner~$W$ measures the spin state~$S$ and the brain state of~$F_2$ in a basis 
   \begin{IEEEeqnarray*}{RL}
     \big\{ &\ket{O} = \sqrt{1/2} ( \ket{\da,D} - \ket{\ua,U} ),\\
     &\ket{F} = \sqrt{1/2} ( \ket{\da,D} + \ket{\ua,U} ) \big\}.
   \end{IEEEeqnarray*}
  \item At time~$t_5$, Wigner and his assistant compare their results. If they both obtain the measurement result $\ok$ corresponding to $\ket{O}$ and $\ket{o}$ respectively, they halt, otherwise the restart with the first step.
\end{enumerate}
\end{protocol}

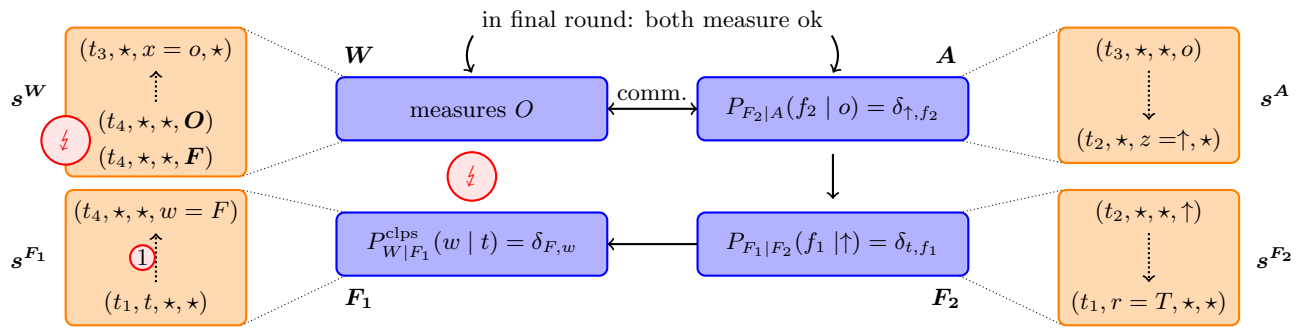
\begin{figure*}
\begin{tikzpicture}[scale=0.6, box/.style={thick, color=blue, fill=blue!30, rounded corners=3pt},
                          sbox/.style={thick, color=orange, fill=orange!30, rounded corners=3pt},
                          box_plot/.style={thick, color=orange, fill=orange!30, rounded corners=3pt},
                          every node/.style={font=\small}]
  \node (comment) at (0,3.5) {in final round: both measure ok};
  \draw [->, thick] (comment.south west) to[bend right] (-4,2.3);
  \draw [->, thick] (comment.south east) to[bend left] (4,2.3);
  \begin{scope}[shift={(-4,1.5)}]
    \draw[ box] (-3.0,-.7) rectangle (3.0,.7);
    \node at (-2.5,1.2) {$\bm{W}$};
    \node(W_e0) at (0,0){measures~$O$};
  \end{scope}
  \begin{scope}[shift={(4,1.5)}]
    \draw[ box] (-3.0,-.7) rectangle (3.0,.7);
    \node at (2.5,1.2) {$\bm{A}$};
    \node (A_e1) at (0,0) {$P_{F_2\vert A}(f_2\mid o) = \delta_{\ua, f_2}$};
  \end{scope}
  \begin{scope}[shift={(4,-1.5)}]
    \draw[ box] (-3.0,-.7) rectangle (3.0,.7);
    \node at (2.5,-1.2) {$\bm{F_2}$};
    \node (F2) at (0,0) {$P_{F_1\vert F_2}(f_1\mid \ua) = \delta_{t, f_1}$};
  \end{scope}
  \begin{scope}[shift={(-4,-1.5)}]
    \draw[ box] (-3.0,-.7) rectangle (3.0,.7);
    \node at (-2.5,-1.2) {$\bm{F_1}$};
    \node (F1) at (0,0) {$P_{W\vert F_1}^{\text{clps}}(w\mid t) = \delta_{F, w}$}; 
  \end{scope}
  \node[draw, thick, color=red, fill=red!10, circle] at (-4,0) {$\Large \lightning $};
  \draw[<->, thick] (-1,1.5) to node[midway, above] {comm.} (1,1.5);
  \draw[->, thick] (4, 0.5) to (4,-0.5);
  \draw[->, thick] (1,-1.5) to (-1,-1.5);
  \begin{scope}
    \begin{scope}[shift={(-11,1.8)}]
      \draw[ sbox] (-2,-1.8) rectangle (2,1.5);
      \node at (-2.8,0.0) {$\bm{s^W}$};
      \node(W_e0) at (0,-0.6){$(t_4,\star,\star,\bm{O})$};
      \node (W_e1) at (0,1) {$(t_3,\star, x=o, \star)$};
      \node (W_e2) at (0,-1.4) {$(t_4,\star, \star,\bm{F})$};
      \node[draw, thick, color=red, fill=red!10, circle] at (-2,-1.0) {$\Large \lightning $};
    \end{scope}
    \begin{scope}[shift={(11,1.8)}]
      \draw[ sbox] (-2,-1.5) rectangle (2,1.5);
      \node at (2.8,0.0) {$\bm{s^A}$};
      \node (A_e1) at (0,1) {$(t_3,\star, \star,o)$};
      \node (A_e2) at (0,-1) {$(t_2,\star, z=\ua,\star)$};
    \end{scope}
    \begin{scope}[shift={(11,-1.8)}]
      \draw[ sbox] (-2,-1.5) rectangle (2,1.5);
      \node at (2.8,-0.0) {$\bm{s^{F_2}}$};
      \node (F2_e1) at (0,1) {$(t_2,\star, \star,\ua)$};
      \node (F2_e2) at (0,-1) {$(t_1, r=T, \star,\star)$};
    \end{scope}
    \begin{scope}[shift={(-11,-1.8)}]
      \draw[ sbox] (-2,-1.5) rectangle (2,1.5);
      \node at (-2.8,-0.0) {$\bm{s^{F_1}}$};
      \node (F1_e1) at (0,1) {$(t_4,\star, \star,w=F)$};
      \node (F1_e2) at (0,-1) {$(t_1, t, \star,\star)$};
    \end{scope}
    \draw[->, thick, densely dotted] (W_e0) to node[midway, left]   {} (W_e1);
        \draw[->, thick, densely dotted] (A_e1) to node[midway, right]  {} (A_e2);
        \draw[->, thick, densely dotted] (F2_e1) to node[midway, right] {} (F2_e2);
        \draw[->, thick, densely dotted] (F1_e2) to node[solid, circle, draw=red, fill=red!10, midway, left, inner sep=0.7]  {1} (F1_e1);
      \end{scope}
  \draw[densely dotted] (6.8,2.2) -- (9.1,3.3);
  \draw[densely dotted] (6.8,0.8) -- (9.1,.3);
  \draw[densely dotted] (6.8,-2.2) -- (9.1,-3.3);
  \draw[densely dotted] (6.8,-0.8) -- (9.1,-.3);
  \draw[densely dotted] (-6.8,2.2) --  (-9.1,3.3);
  \draw[densely dotted] (-6.8,0.8) --  (-9.1,.0);
  \draw[densely dotted] (-6.8,-2.2) -- (-9.1,-3.3);
  \draw[densely dotted] (-6.8,-0.8) -- (-9.1,-.3);
\end{tikzpicture}
\caption{ Structure of the contradiction for the setup in~\cite{FrRen}: Each party \emph{deduces}, based on their own result, the measurement outcome of another party with certainty (dotted arrows). \emph{Compatibility constraints} then require this deduction to equal the \emph{observation} of the respective party (solid arrows). The deduction (1) of~$F_1$ is, however, a result of the subjective-collapse model.}
\label{fig:setup_extWigner}
\end{figure*}

\section{Discussion of the Subjective-Collapse Model}
\noindent
We will now consider the predictions of the various agents about the other's measurement result in the halting round of the protocol.
The calculations are performed in the Everett formalism described above.
The subscript on the probabilities indicates whether a collapse was considered, $P_{\text{clps}}$, or not, $P_{\text{ism}}$.
\paragraph{$A$ about~$F_2$} The knowledge of~$A$ about~$F_2$ is given by the distribution
\begin{IEEEeqnarray*}{RL}
  &P_{\text{ism}}(f_2\mid a) = \\ 
  &1/N_a \tr\left[ V \proj{\phi_C} V^{\dagger} \cdot \proj{f_2} \otimes \proj{a}  \otimes \id_{\text{rest}}\right]
\end{IEEEeqnarray*}
where $V = V_A \cdot V_{F_1} \cdot V_{F_2}$
with the values
\begin{equation}\label{eq:condProbF2}
  \begin{array}{c|c|c}
   f_2 & P_{\text{ism}}( f_2\mid f) & P_{\text{ism}}( f_2\mid o) \\ \hline
   u & 0.2 &  1.0  \\
   d & 0.8 &  0.0 
  \end{array}
\end{equation}
Thus,~$A$ can conclude that~$F_2$ measures $\ua$ upon obtaining~$o$.
\paragraph{$F_2$ about~$F_1$} The knowledge of~$F_2$ about~$F_1$ is given by the distribution
\begin{IEEEeqnarray*}{RL}
  P_{\text{ism}}(f_1\mid f_2) =& 1/N_{f_2} \cdot \tr\big[ V \proj{\phi_C} V^{\dagger} \cdot \\
  &\proj{f_1} \otimes \proj{f_2}  \otimes \id_{\text{rest}}\big]
  \end{IEEEeqnarray*}
where $V = V_{F_1} \cdot V_{F_2}$
with the values
\begin{equation}\label{eq:condProbF2}
  \begin{array}{c|c|c}
   f_1 & P_{\text{ism}}( f_1\mid U) & P_{\text{ism}}( f_1\mid D)  \\ \hline
   h &  0.0 & 0.5 \\
   t &  1.0 & 0.5 
  \end{array}
\end{equation}
Then~$F_2$ can conclude that~$F_1$ measures~$t$ upon measuring~$\ua$.
\paragraph{$F_1$ about~$W$ (collapse)} If~$F_1$ assumes a collapse to happen after his measurement he can calculate the conditional probability over~$W$'s result as follows.
\begin{IEEEeqnarray*}{RL}
  P_{\text{clps}}(w\mid f_1) =&1/N_{f_1} \cdot  \tr\big[ V \proj{f_1} V^{\dagger} \cdot \\
  &\id_{\text{rest}} \otimes \proj{w} \big]
\end{IEEEeqnarray*}
where $V= V_W\cdot V_A\cdot V_{F_2} \cdot V_{F_1} $.
This yields the distribution
\begin{equation}\label{eq:condProbF1}
  \begin{array}{c|c|c}
   w & P_{\text{clps}}( w\mid H) & P_{\text{clps}}( w\mid T)  \\ \hline
F & 0.5 &  1.0 \\
O & 0.5 &  0.0 
  \end{array}
\end{equation}
So upon measuring~$T$, $F_1$ can conclude that~$W$ measured~$F$.
\paragraph{$F_1$ about~$W$ (no collapse)} The distribution turns out to be 
\begin{IEEEeqnarray*}{RL}
  P_{\text{ism}}(w\mid f_1) =& 1/N_{f_1} \cdot \tr\big[ V \proj{\phi_C} V^{\dagger} \cdot \\ 
  &\proj{f_1} \otimes \proj{w}  \otimes \id_{\text{rest}}\big]
\end{IEEEeqnarray*}
with values
\begin{equation}\label{eq:condProbF1}
  \begin{array}{c|c|c}
   w & P_{\text{ism}}( w\mid H) & P_{\text{ism}}( w\mid T)  \\ \hline
   F & 0.83333 & 0.83333\\
   O & 0.16667 & 0.16667
  \end{array}
\end{equation}
if there is no collapse. Hence~$F_1$ cannot deduce~$W$'s measurement with certainty.
The contradiction arises from~$F_1$'s prediction about~$W$'s result in the halting round as shown in the Fig~\ref{fig:setup_extWigner}. 
This prediction is a result of~$F_1$ attributing a collapse to his measurement and hence due to the subjective-collapse model. 
In the halting round, if~$W$ actually measured~$O$, and~$F_1$ claimed he should have obtained another result, the model~$F_1$ used to derive the very claim is falsified.
\emph{In general, the subjective-collapse model does not give consistent predictions for a setup containing encapsulated observers.}

 \section{Conclusion}
\noindent
In this note we critically review a recent result~\cite{FrRen} by Frauchiger and Renner questioning the logical consistency of single-world interpretations of quantum theory.

In fact, the authors of \cite{FrRen} prove an inconsistency of the subjective-collapse framework.
It shows that \emph{collapse and unitary evolution} cannot be reconciled. 
This is the core of the \emph{``big'' measurement problem}~\cite{brukner2015quantum} as illustrated in the original Wigner's-friend experiment.
The contradiction in the latter is based on the possibility of additional measurement results, and can thus merely be tested in a statistical analysis of multiple runs.
In the Frauchiger/Renner protocol, however, two parties make opposing predictions for a \emph{single} measurement result (once the halting condition is met):
Their protocol makes the conflict more striking and evident.

The contradiction cannot be resolved with the notion of ``many worlds'' established in \cite{FrRen} without abandoning the definiteness of the measurement result. 
This, in turn, would be to say one does not infer distinguishable information from a measurement. 
We consider $\AND$-related statements about results of \emph{a single measurement for a given observer} debatable, regardless of the interpretation of quantum theory. 
For instance, Everett's formalism does not imply multiple ``coexisting'' results in the sense above.
One of the branches in common many-worlds interpretation has to be singled out if measurement results are to be definite (see Fig~\ref{fig:many worlds}).
An observer merely holds his memory system, and the respective density matrix is obtained by partial trace over the source. This yields a statistical mixture of memory states corresponding to definite outcomes instead of the pure state as claimed, for example, in \cite{Maudlin95}.  
This point is connected to Maudlin's taking an \emph{external} standpoint of a superobserver, viewing the wave function from the outside.
For any \emph{specific} (internal) observer, measurement results keep being definite.
In this sense, definiteness of outcomes can be incorporated into the relative-state formalism and is not necessarily equivalent to an objective collapse.
Subjectively, a measurement \emph{looks like} a collapse as an observer cannot ``watch'' himself performing the observation --- and thus cannot ``see'' the unitary evolution entangling him with the observed system \cite{filan2015would}. 

The formal description of the subjective collapse leads to contradicting predictions for Wigner's-friend experiments. 
The conflict becomes manifest only if encapsulated observers can communicate --- i.e., if the friend and Wigner can test their predictions against actual observations. 
It has been questioned whether this is in principle possible.

The Wigner's-friend experiment can (in principle) discriminate between two competing quantum formalisms describing a measurement --- the unitary relative-state formalism and the non-unitary measurement update rule. 
A specific combination of these two formalisms, together with the assumption regarding possible communication, gives a contradiction.
We do, however, not regard a \emph{formalism} to necessarily imply a particular \emph{interpretation} like ``many worlds'' or ``collapse.''
We believe that the contradiction above does, therefore, not disqualify a particular interpretation of quantum mechanics.

\begin{acknowledgments}
\noindent
This work is supported by the Swiss National Science Foundation (SNF), the \emph{NCCR QSIT}, and the \emph{Hasler Foundation}. We would like to thank \"Amin Baumeler, Boyan Beronov, Paul Erker, \v Caslav Brukner, and Renato Renner for helpful discussions.
\end{acknowledgments}

\end{document}